\documentclass[aps,prd,amsmath,nofootinbib,floatfix,fleqn]{revtex4}
\pdfoutput=1 
\usepackage{color}
\usepackage[dvipsnames]{xcolor}
\usepackage{graphicx}
\usepackage{bm}
\usepackage[utf8]{inputenc}
\usepackage{subfig}
\usepackage{pythonhighlight}
\usepackage{amsfonts}
\usepackage{float}
\usepackage{epsfig}
\usepackage{pdfpages}
\usepackage{wrapfig}

\definecolor{darkblue}{rgb}{0.0,0,0.5}
\definecolor{darkgreen}{rgb}{0.0,0.3,0.0}
\definecolor{redish}{rgb}{0.675,0,0.2}
\definecolor{red}{rgb}{0.8,0,0}
\definecolor{green}{rgb}{0,0.6,0}
\definecolor{blue}{rgb}{0,0,0.8}

\usepackage[unicode=true, bookmarks=false, linkcolor = darkblue, citecolor = redish, breaklinks=false, colorlinks=true, hyperfootnotes=true]{hyperref}

\newcommand{\ren}{R\'enyi }

\begin{document}

\date{\today}
\title{Information Criteria for Selecting Parton Distribution Function Solutions}

\preprint{}

\author{Aurore Courtoy$^{(a,1)}$, Arturo Ibsen$^{(a,2)}$}

\affiliation{$^{(a)}$Instituto de F\'isica, Universidad Nacional Aut\'onoma de M\'exico, Apartado Postal 20-364, 01000 Ciudad de M\'exico, M\'exico }
\email{$^{1}$aurorecourtoy@gmail.com, $^{2}$a.i.v.r@ciencias.unam.mx}

 \date{\today}

\begin{abstract}
In data-driven determination of Parton Distribution Functions (PDFs) in global QCD analyses,  uncovering the true underlying distributions is complicated  by a highly convoluted inverse problem. 
The determination of PDFs can be understood as the inference of a function supported on $[0,1]$, a problem that admits multiple acceptable solutions. An ensemble of solutions exists that pass all standard goodness-of-fit criteria. 

In this paper, we propose algorithms for the classification, clustering, and selection of solutions to the determination of PDFs, or any functions on $[0,1]$, based on the characterization of their shape. We explore information-theoretic based (\ren entropy and divergence) and optimal-transport based (Wasserstein distance) criteria. In particular, we advocate for the use of the \ren entropy as an {\it absolute} estimator per solution, as opposed to {\it relative} estimators that compare solutions  pairwise. We show that the \ren entropy can  characterize the space of solutions {\it w.r.t.} the PDF shapes. Paired with the identification of the optimal combination of solutions via Pareto fronts, it provides a plausible and minimalist selection algorithm. Moreover, \ren entropy proves versatile for use in clustering applications.
\end{abstract}

\maketitle

\section{Introduction}

Distribution functions play a central role in scattering processes involving hadrons in the deep inelastic regime, where  cross sections can be factorized into hard and soft contributions. These non-perturbative mathematical objects, the PDFs,  depend on two variables, the momentum fraction $x$ and the hard scale $Q^2$. While the behavior of PDFs with respect to the latter is known from perturbative QCD, their shape as functions of $x$ is not. This characteristic is inherent to low-energy dynamics and thus largely inaccessible from first principles, although many efforts have been put forth in recent years, {\it e.g.} by the Schwinger-Dyson or, more importantly, the lattice QCD communities~\cite{Lin:2017snn,Constantinou:2020hdm}. 

PDFs are more reliably determined through data-driven analyses in the QCD framework,   
where they are extracted  by solving  an inverse problem. 
The state-of-the-art proton PDFs~\cite{Hou:2019efy,Bailey:2020ooq, NNPDF:2021njg} result from  analyses that combine experimental, theoretical and statistical tools and frameworks, aiming to tackle the many aspects of the PDF inverse problem. 
 Progress in PDF inference  has reached a unprecedented refinement in terms of statistical treatment, in particular, when it comes to quantifying the uncertainties on PDFs, see {\it e.g.}~\cite{Courtoy:2022ocu, Kotz:2023pbu, Kotz:2025lio, Kotz:2025une,Harland-Lang:2024kvt, Costantini:2024wby}.

The  inverse problem, formulated through optimization of an objective function, consists in inferring models for PDFs. 
The convolutions inherent to the theoretical description of observables make the resulting inverse problem ill-posed, {\it i.e.} it does not admit a unique solution.
A model for PDF consists of a  mathematical basis and a set of hyperparameters, each  leading to a central solution accompanied by an Hessian matrix or, in the Monte Carlo approach, to a distribution of solutions -- both reflecting  the  propagation of the experimental ({\it aleatoric}) uncertainty.
Commonly adopted approaches  model PDFs using a single, so-called ``fixed," functional form (a parametric approach) or by neural networks (a non-parametric approach).  A recently developed non-parametric method instead  generates polynomials by varying hyperparameters~\cite{Kotz:2025une}.
As most non-parametric approaches produce ensembles of solutions, 
dedicated tools are required to classify them.
This is particularly relevant for addressing parametrization bias or model uncertainty. The non-uniqueness of the plausible models  manifests itself through the coexistence of multiple solutions,  functions of $x$, {\it i.e.} the PDF shapes. How this ensemble of solutions is accounted for contributes to the global analysis's methodology; for the CTEQ-TEA group,  it gives rise to an {\it epistemic} uncertainty~\cite{Courtoy:2022ocu}. \\

In this context, a prescription for model combination was designed~\cite{Gao:2013bia, Kotz:2025lio}, which raises the following question: How can we select the most diverse shapes that span the space of solutions? Addressing this question requires introducing estimators, such as metrics or information criteria, to characterize the various solutions as shapes, across various flavor combinations. Specifically, we propose using the \ren entropy, a generalization of the Shannon entropy, to quantify shapes in terms of their {\it bumpiness} or {\it flatness}. As we will argue, the \ren entropy is an {\it absolute}, as opposed to {\it relative}, quantity: it quantifies each curve independently, without reference to another solution or curve. It is also combinatorially more efficient than evaluating relative quantities, which will also be discussed and compared to. Once the \ren space defined, the most representative shapes can be selected. To this end, while there are many algorithms available to cluster and characterize such optimal solutions, we employ the concept of Pareto fronts, developed in the context of multiobjective optimization.

We illustrate how \ren entropy can be used to  select, classify, and  cluster families of shapes on $x\in [0,1]$, using the PDFs of the pion generated within the Fant\^omas framework~\cite{Kotz:2023pbu}. For the FantoPDFs, the selection of the most diverse shapes, called {\sf metamorphs}, was originally performed ``by hand." The present manuscript upgrades this feature, which could be extended to families of PDFs with more flavor combinations, effectively using the pion PDF analyses as a sandbox for proton studies.\\

The manuscript is organized as follows.  Section~\ref{sec:pdfs} outlines the motivation for classifying and selecting solutions of PDFs. The core of the paper consists in the selection criteria, distinguishing between {\it absolute} and {\it relative} types. 
We show that the results obtained through the \ren-entropy based algorithm (Section~\ref{sec:absolute}) are qualitatively comparable to using {\it relative} criteria, such as the \ren divergence or the Wasserstein distance~(Section~\ref{sec:relative}).  Comments and caveats are discussed in Section~\ref{sec.discussion}, and further in the conclusions (Section~\ref{sec:conclusions}). Mathematical details,  specifically of the Pareto fronts,  are provided in the Appendices~\ref{app:tweakGauss}-\ref{app:pareto}.

\section{Parton Distribution Functions}
\label{sec:pdfs}

Parton Distribution Functions characterize the behavior of quarks and gluons as a function of the momentum fraction $x$ of the parent hadron they carry when interacting with a hard probe, at a factorization scale $\mu$. PDFs are mathematical objects, formally defined through factorization theorems, with a formulation on the light cone. For example, the quark PDFs $f_{a/p}$ representing the distribution of quarks of flavor $a$ in a hadron $p$ are defined in the $\overline{\rm MS}$ factorization scheme as  
\begin{equation}
\label{eq:MSbarPDF}
f_{a/p}(x,\mu^2)=\frac{1}{4\pi} \int dz^- e^{-i x P^+z^-}
\left\langle P|\bar{\psi}_a(0,z^-, {\bf 0}) \gamma^+ W(z^-,0)\psi_a(0)|P\right\rangle
\;,
\end{equation}
where we have used the light-cone coordinates, $a^{\pm}=(a^0\pm a^3)/\sqrt{2}$ and $a^{\mu}=(a^+, \vec{a}_{\perp}, a^-)$, and
\begin{equation}
\label{eq:WilsonLine}
W(z^-,0) = {\cal P} \exp\left(
-i g \int_0^{z^-}\!\!d\bar z^-  \widehat A^+(0^+,\bar z^-,\vec{0}_T)
\right)
\end{equation}
is the Wilson eikonal line that ensures color gauge invariance. In the light-cone gauge, and to LO, PDFs have been interpreted as Probability Density Functions, which is intuitively the distribution of the momentum fraction $x$ among all members, $a \in \{u, \bar{u}, d, \bar{d}, s, \bar{s}, g, \cdots\}$ a flavor that contributes to 
the proton's structure and $a \in \{u, \bar{d}, \bar{u}=d=s=\bar{s},  g\}$ a flavor that contributes to  
the pion's structure. Beyond those approximations, PDFs do no longer follow a probability density interpretation. Still, their support is $x\in [0,1]$ for quarks and gluons ; they are derivable on $x\in ]0,1[$ and serve as weights to the $x$'s $n$ projections, called Mellin moments.  These properties lead to the integrability rules
\begin{eqnarray}
    \int_0^1\, dx \, x^{n>0}\, f_{\{S, g\}/p}(x, Q^2) &&{\rm \quad are \;finite}\, ,\nonumber \\
    \int_0^1\, dx \, x^{n\geq 0}\, f_{v/p}(x, Q^2) &&{\rm \quad are \; finite}\, ,
    \label{eq:integrability}
\end{eqnarray}
where $v$ refers to  valence, and $S, g$ to sea and gluon, respectively.
The integrability of PDF's projections builds on the Operator Product Expansion connection to PDFs and the relation to Mellin moments as defined through local operators. Some of those moments reflect conservation laws through  sum rules, {\it e.g.} the valence sum rule corresponding to $n=0$, and are hierarchically important {\it first-principle} constraints on PDFs. 

Besides the above-mentioned properties, the shape of PDFs is largely unconstrained by the theory of the strong interaction itself, as PDFs appear to be non-perturbative object at the core -- they bridge the degrees of freedom of perturbative QCD to bound-states. Hence, the determination of PDFs as functions of $x$ is truly twofold -- uncovering the low-energy dynamics that binds partons into hadrons, and using solely a perturbative framework for data-driven analyses to infer such shapes. 

The inference of PDFs results of global QCD analyses of large pools of data with an extended coverage in $(x, Q^2)$, where $Q^2$ is often identified with $\mu^2$. Each such optimization problem aims to find the family of functions, $f_{a/p}$ with $a$ component of  vector, which length and composition depends both on the flavor structure of the hadron and the experimental access to processes that are sensitive to those flavors. For example, while proton PDFs span spaces that can be as large as 9-dimensional, the pion PDFs are experimentally less constrained and a 3-dimensional space is the best that can be explored at this time. Of relevance to the present analysis, we will focus on  $n_{\rm flavor}$-dimensional spaces in which the families of functions are defined. 

To each direction of the $n_{\rm flavor}$-dimensional space corresponds a function of $x$, which, more often than not, is given in terms of polynomials, characterized by its generating function (or basis), its hyperparameters, and its vector of (effective) free parameters. The latter are commonly used to characterize the dimensions of the problem. For a single-family solution,  they are best illustrated by the PDFs obtained in the Hessian formalism. For example, for the pion PDFs, the xFitter group used a fixed parametrization with $4 (v)+3 (S)+2(g)=9$ parameters~\cite{Novikov:2020snp}, corresponding to a 9-dimensional eigenvector space for the Hessian matrix, and the objective function. On the other hand, a baseline solution for the proton PDF spans, at least, a 26-dimensional parameter space~\cite{Hou:2019efy}.\\

Heavily convoluted inverse problems do not have a unique solution, and, as such, allow for multiple shapes of PDFs. While not all solutions are statistically acceptable, {\it w.r.t.} goodness-of-fit and analytical properties mentioned above, many constitute very good complementary solutions. 
 Once filtered through $\chi^2+\delta \chi^2$ and physical priors (such as soft positivity bounds), those are qualitatively equally good~\cite{Kovarik:2019xvh}.
They are expected to reflect the imperfect kinematical coverage of the $(x, Q^2)$ plane. As such, the various acceptable solutions should span the extrapolation regions to result in large uncertainties, {\it i.e.} the uncertainty is large where no data or no theoretical penalties are constraining the shapes.

Accounting for more families of solutions means that diverse models  are generated and tried against the data. 
Models for PDFs are built on a basis ${\cal Q}$
\begin{eqnarray}
    x\, f_{a/h}(x, Q_0^2)&=& A_a\, x^{\alpha_a} (1-x)^{\beta_a}\times {\cal Q}(g[x]; \underline{c}, \underline{c}^h)\, ,
\end{eqnarray}
where $\underline{c}^h$ is the vector of hyperparameters inherent to the model, and the optimization determines the best values for the vector of free parameters, $\{A_a, \alpha_a, \beta_a; \underline{c}\}$. 
There exist various proposals to generate ensembles of models to sample from. We will mainly focus on the non-parametric, Fant\^omas methodology~\cite{Kotz:2025une}.  To generate {\sf metamorphs}, ${\cal Q}=1+{\cal B}$, with ${\cal B}$ a B\'ezier curve, and $\underline{c}^h=\{N_m, \underline{x}, g[x]\}$, respectively, the degree of the polynomial, the vector of abscissa for the control points, and the scaling function of the argument. 
Each set of $\underline{c}^h$ generates a different model, all of which are not necessarily related one to another.\footnote{The B\'ezier curve formalism allows to relate polynomials of different degrees easily.} 
A family of solutions can thus be represented as $\{x\, f^k_{v/\pi}(x, Q_0^2),\, x\, f^k_{S/\pi}(x, Q_0^2),\, x\, f^k_{g/\pi}(x, Q_0^2)\}$ for the pion PDFs, with $k\in {1,..., n_s}$ labeling the $k^{th}$ of all  $N_s$ acceptable solutions. 
\\

The classification of solutions is reduced to the characterization of that $n_{\rm flavor}$-dimensional space.
Conceptually, we switch the focus from parameter space to an $n_{\rm flavor}$-dimensional space. In doing so, we acknowledge the difficulty of mapping the free parameters between the different models. In that regard, it is akin to the philosophy underlying the use of neural networks for PDF determination. This peculiarity renders the cross-analysis of the families of solutions rather intricate. To streamline this analysis, we conceptually integrate the free parameters of each flavor into the shape of the PDFs. The latter then need to be characterized, 
which we do through an  elegant strategy, as described below.
. 

\section{Selection criteria}

\begin{figure}[t]
     \centering
     \includegraphics[width=0.45\linewidth]{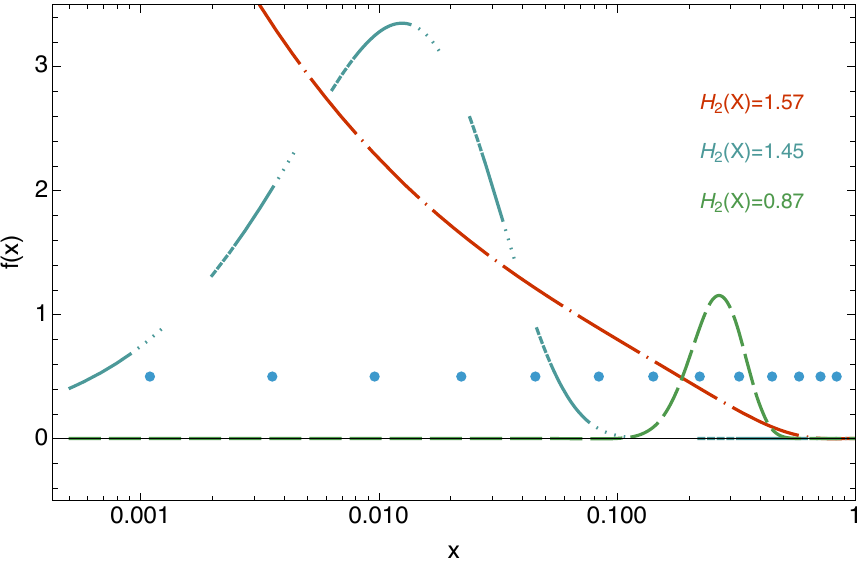}
     \includegraphics[width=0.45\linewidth]{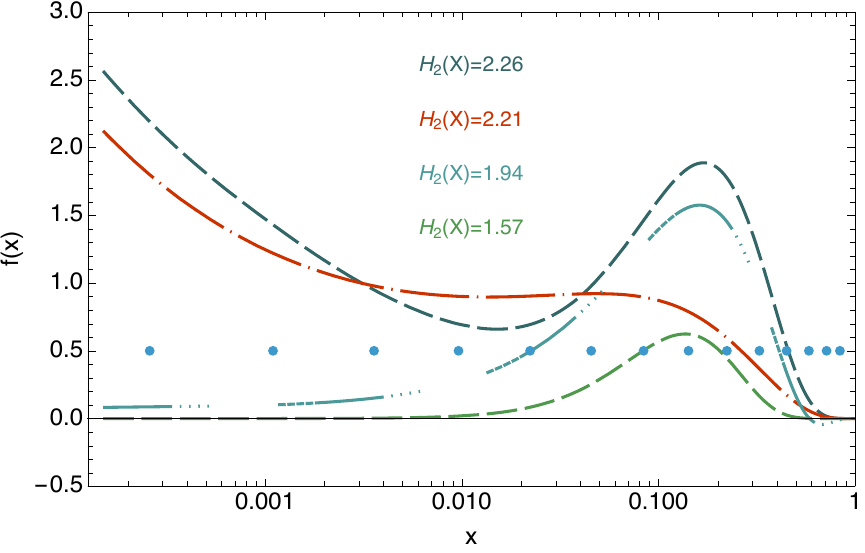}
\caption{  Illustration of values for $H_2(X)$ for various shapes. (Left) ideal case of shapes ranging from (in decreasing order) steep (dot-dashed red),  spread-out (3-dot-dashed cyan) and highly-concentrated (dashed green), for tweaked Gaussian point (blue dots). (Right) similar color code and characteristics, with the addition of the short-dashed blue-grey curve whose $H_2(X)$ is similar to the steep curve. Both show a similar pattern in variations, not magnitude, at the evaluation points.}
\label{fig:concentration}
\end{figure}

The sets of solutions described above must hence be classified or organized to obtain a full span of the space of solutions. 
In the case of sampling over models,  the parametrizations, or models, that led to the solutions do not emerge from an obvious probability distribution. Hence there should be no notion of density in how the space is populated. 
On the contrary, the entire set of solutions illustrates the tossing of markers into the hyperparameter space, each toss generating a model.  The resulting models can cluster in neighboring regions.
As such, keeping only those few solutions that best characterize the space  is the task that we will be describing here below.  Our main results are found using {\it absolute shape estimators}, Section~\ref{sec:absolute}. They are challenged and confirmed using {\it relative criteria}, such as distances and divergences~\ref{sec:relative}.

\subsection{Absolute  criterion}
\label{sec:absolute}

Models for PDFs consist in functional forms of the variable $x\in [0,1]$ and a vector of training parameters, $\underline{c}'$, while each model is a realization of a basis cast into a functional form through a choice of hyperparameters, $\underline{c}^h$. What characterizes the various solutions is the shape of the solution, {\it e.g.} bumpy, flat, exponentially falling off, etc. An absolute description of each curve can be obtained through the evaluation of its R\'enyi entropy, $H_{\alpha}$. Defined for probability densities, $H_{\alpha}$ reflects how certain (uncertain) a random variable (here, the momentum fraction $x$) is. In other words, a pronounced maximum --a bump-- is certain and has a low entropy, while a flat region is uncertain and corresponds to a higher entropy. Even though PDFs are not proper probability densities\footnote{Not all PDFs obey $f_{a/q}(x\to 0)\to 0$.}, using $H_{\alpha}$ as an estimator is still a valid shape characterizer or bump-width filter. 

The \ren  entropy is defined as
\begin{eqnarray}
    H_\alpha(X) = \frac{1}{1 - \alpha} \ln \left( \sum_{i=1}^{n} p(x_i)^\alpha \right), \quad \alpha \neq 1
    \label{eq:renyi}
\end{eqnarray}
where $p(x)$ is the probability density, {\it i.e.}
a discrete random variable \(X\) with possible outcomes \(\{x_1, x_2, \ldots, x_n\}\) and corresponding probabilities \(\{p_1, p_2, \ldots, p_n\}\). The value of $\alpha$ gives emphasis on core or tails, with $\alpha=2$ commonly called the collision entropy or the \ren entropy. Higher values of $\alpha$ will weigh maxima, or bumps, more and the other way round, with the particular case of recovering the Shannon entropy for $\alpha=1$. The choice $\alpha=2$, which we will adopt from now on,  constitutes a good balance between identifying the bumps and accounting for more flat ranges. \\

\begin{figure}[t]
     \centering
     \includegraphics[width=0.45\linewidth]{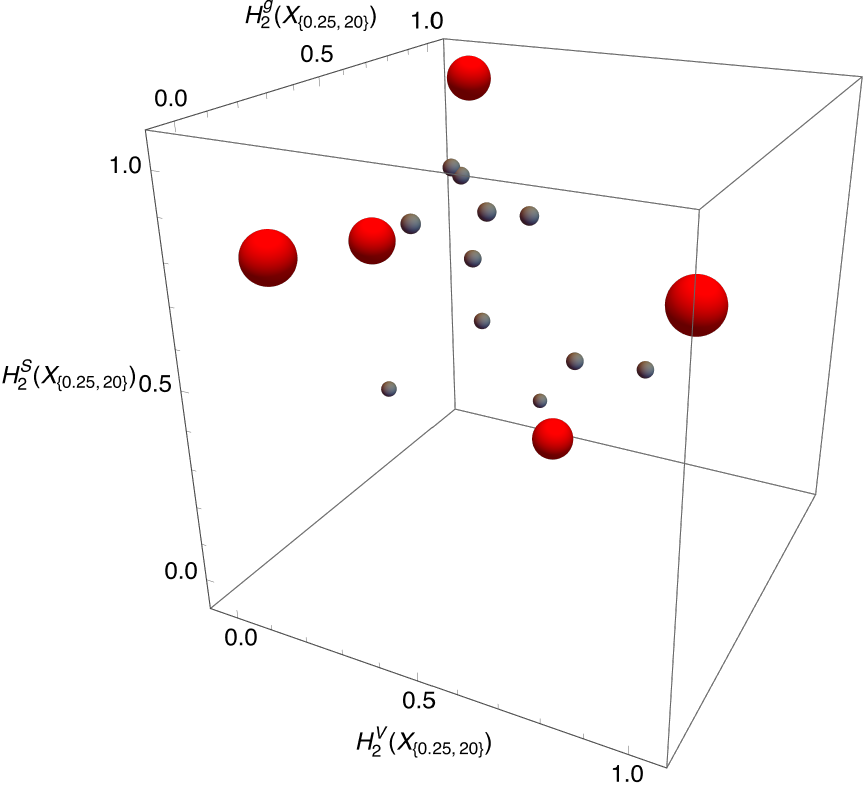}
     \includegraphics[width=0.45\linewidth]{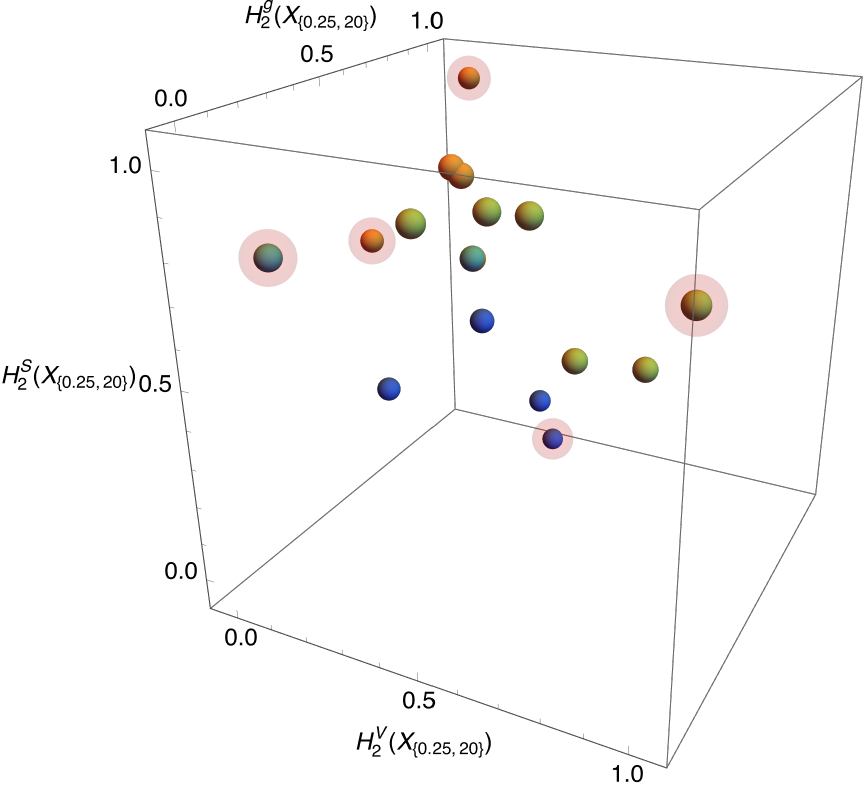}
\caption{R\'enyi 3-dimensional space. Left panel: The light gray points represent all acceptable solutions. Larger red points represent the vertices of the Pareto hypersurfaces, {\it i.e.} floor and roof. Right panel: The four clusters identified through the UMAP nearest-neighbor algorithm (orange, green, cyan, and dark blue), along with the vertices of the Pareto fronts (low opacity red circles).
Both are illustrated for   $X_{0.25, 20}$. }
\label{fig:abs_renyi_3fl}
\end{figure}

For densities normalized according to $f(x_i)/(\sum_i^n f(x_i))$ with $f(x)$ a function akin to PDFs, $H_2(X)$ is bounded by $0$, for a delta function, and $\ln{n}$, for a flat distribution.
For unimodal functions, {\it smooth} enough function at $X$ or function with steep rise (yet integrable) at end-points,
the entropy may vary continuously from highly concentrated, to spread-out, to steep,  and to flat solutions. This is illustrated, in a linear-log scale, in Fig.~\ref{fig:concentration}. The concept of concentration does not reflect the magnitude of the curves, nor the position of the bumps. That is, the most extreme entropies do not necessarily encompass all the ordinates of intermediate solutions.  \\

The evaluation of the \ren entropy is defined for discrete points, $x_i$, and will depend on the $x$-grid that we choose, as it is the case for many estimators. In the case of PDFs, the $x$ spacing can be chosen so to span (or not) the data coverage. We have chosen to use tweaked Gaussian points (see Appendix~\ref{app:tweakGauss}), so that we have a direct access to an integral or kernel interpretation of the argument of the logarithm in Eq.~(\ref{eq:renyi}). In practice, we evaluate
\begin{eqnarray}
    H_2^q(X_{\alpha_x, n, \tau}) = -\, \ln \left( \sum_{i=1}^{n} \left(\frac{x_i\, f_{a/h}(x_i)}{\sum_{j=1}^{n}x_j\, f_{a/h}(x_j)}\right)^2 \right), 
    \label{eq:renyi_pract}
\end{eqnarray}
with the PDFs evaluated at $Q_0$ and $X_{\alpha_x, n, \tau}$ is a truncated ($\tau$) vector of $n$ $x_i$ defined with a scaling $\alpha_x$. We chose to use the $x\, f(x)$ product as it highlights features of the solutions that would otherwise be lost ; it is due to the fact that the integrability of the sea and the gluon is ensured starting from the definition of the average momentum fraction, Eq.~(\ref{eq:integrability}).

A function is neither uniquely nor unambiguously 
determined by its entropy. For example, on the {\it r.h.s} of Fig.~\ref{fig:concentration}, two curves which small-$x$ behavior is similar but differs in the large-$x$ region display a similar value for $H_2(X)$. This is understood by the trend in the curves's derivatives. \ren entropy can nonetheless lead to a satisfactory classification of functions on $[0,1]$.\\

Evaluating the \ren entropy for all families of solutions is straightforward and fast. Once obtained, the values of the \ren entropy can be visualized as coordinates in the $n_{\rm flavor}$-dimensional space of \ren entropy, {\it e.g.} a 3-dimensional space formed by the values of the entropy for the valence, sea and gluon   PDFs of the pion. It is more informative to work with normalized values of $H_2(X)$, so all spaces will be unit cubes, as illustrated in Fig.~\ref{fig:abs_renyi_3fl}. The convex hull of the points can be used to define the volume occupied by the solutions, see {\it e.g.}~\cite{Anwar:2019wrj}. Identifying the most extreme points of that volume would allow to balance the diversity of shapes among all flavors. 
Instead, we choose to analyze Pareto fronts, which give us the floor (and, conversely, the roof)\footnote{In economy, the Pareto floor and roof are referred to as, correspondingly, Pareto front and worse front.} of enveloping solutions,
\begin{eqnarray}
    P_{\rm  floor }(Y)&=& \{ \underline{y}\in Y:\{\underline{y}'\in Y: \underline{y}'\preceq
    \, \underline{y}, \underline{y}\neq \underline{y}'\}\}\nonumber\\
    P_{\rm  roof}(Y)&=& \{ \underline{y}\in Y:\{\underline{y}'\in Y: \underline{y}'\succeq
    \, \underline{y}, \underline{y}\neq \underline{y}'\}\}
    \label{eq:pareto}
\end{eqnarray}
with $Y\equiv H_{2, X_{\alpha_x, n, \tau}}^q$ and $\underline{y}$ correspond to the  coordinates, {\it e.g.} $\{ H_2^V, H_2^S, H_2^g\}$ for the pion PDF. The symbols $\{\preceq, \succeq\}$  denote the componentwise partial order, and are defined in Appendix~\ref{app:pareto}.

\begin{wrapfigure}{r}{0.5\textwidth}
 \centering
\includegraphics[width=0.9\linewidth]{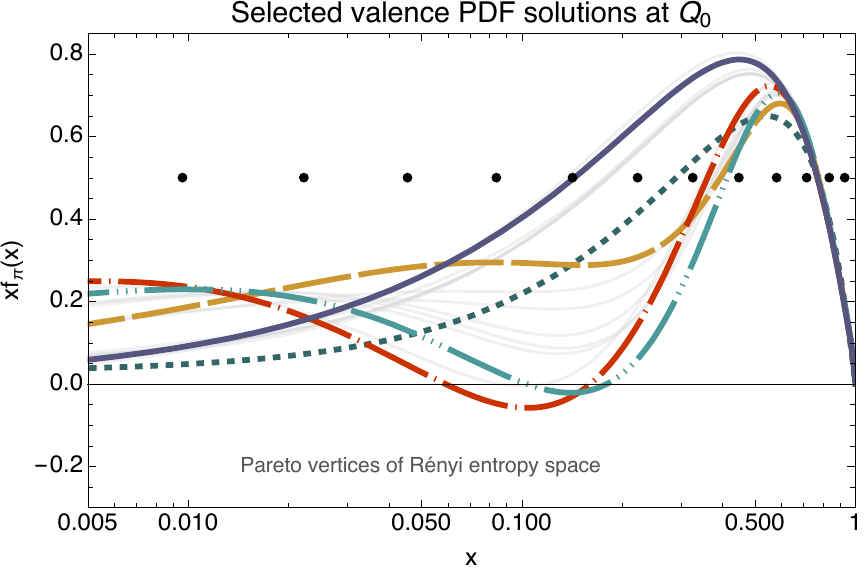}
\includegraphics[width=0.9\linewidth]{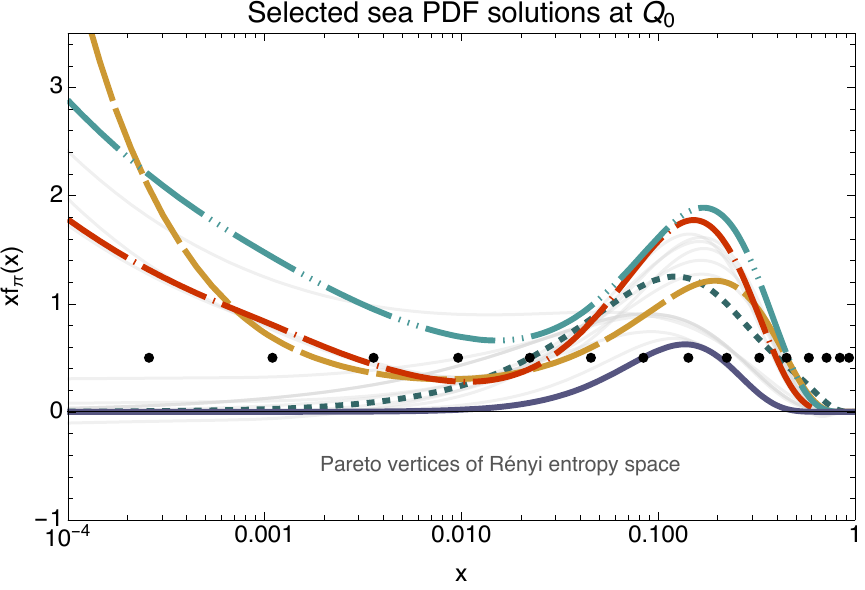}
 \includegraphics[width=0.9\linewidth]{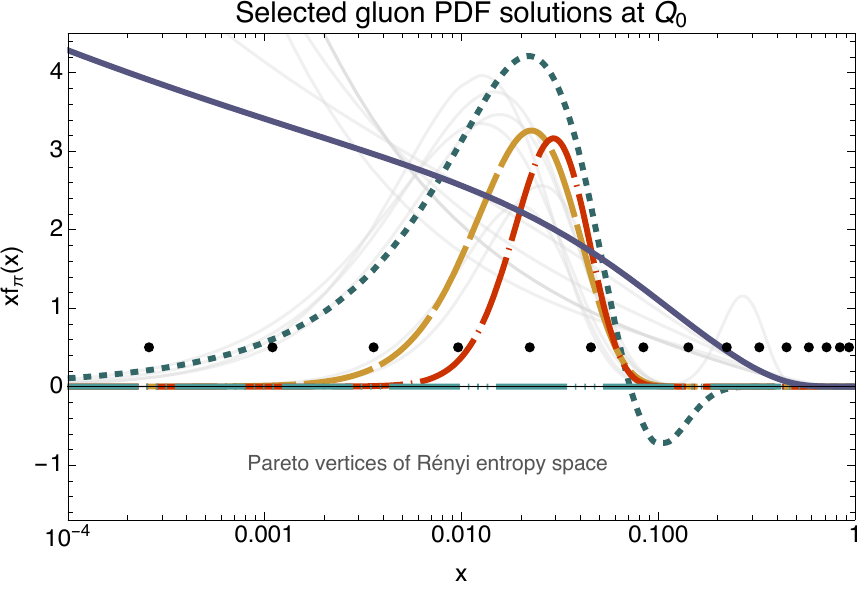}
\caption{Selected shapes (color) of the pool of acceptable metamorph solutions generated for  the Fant\^omas pion PDFs (light grey). R\'enyi entropy-based coordinates are filtered as Pareto vertices. The black dots represent the $x$ values at which the entropy was evaluated.}
\label{fig:abs_renyi_vertices}
\end{wrapfigure}

The Pareto fronts represent the optimal balance between all $n_{\rm flavor}$ coordinates in either a minimal or maximal fashion. While all PDFs are related via the momentum sum rule, {\it e.g.} there is already a balance in the momentum fractions, the tradeoff revealed  by the Pareto fronts arises in identifying the diversity in entropies across flavors. 
For a point on the Pareto floor, if one coordinate is large, at least one of the other two must be smaller; similarly, for the Pareto roof, if one coordinate is small, at least one other must be larger. This balance is not guaranteed for points in between the fronts.  
The obtained Pareto fronts, while discrete, can be seen as approximating continuous hypersurfaces derived through ordering. This allows the approach to be extended to flavor spaces with $n_{\rm flavor} > 3$, as illustrated here, providing a visually intuitive picture.

Since the fronts can still be large ensembles, we study the effect of selecting a subset, corresponding to the extrema of both fronts in all $n_{\rm flavor}$ directions,
\begin{eqnarray}
    &&\underline{y}_{\min} \preceq \underline{y}' \preceq \underline{y}_{\max}, \\\nonumber\\
    &&{\rm with}\nonumber\\
    &&
    \underline{y}_{\min/\max}=\underset{\underline{y} \in P_{\mathrm{floor}}(Y) \cup P_{\mathrm{roof}}(Y)} {\min/\max} y_i\quad \forall i=1,\cdots, n_{\rm flavor}\quad.\nonumber
\end{eqnarray}

The vertices of the Pareto set, taken as the corner of the space, correspond to specific solutions for the PDFs. 
Given that PDFs are {\it smooth} enough, defining intervals of extrema of the \ren entropy, as defined in Eq.~ (\ref{eq:renyi_pract}), is sufficient to encompass most bounding shapes\footnote{See Appendix~\ref{app:diffGlue} for discussion of exceptions.}.

In the Fant\^omas analysis, we tested  about 100 solutions,   of which we retain only $N_s\sim 20$ that pass both the $\chi^2+\delta\chi^2=450$ threshold and  a soft positivity constraint\footnote{A similar analysis without the soft constraint on positivity, retaining about $\sim 40$ solutions, leads to similar results.}.
By applying the above-described methodology to the $N_s$ solutions, we obtained 5 vertices that characterize the extrema of the three-dimensional space formed by the value of R\'enyi entropy for each flavor. They are shown in red on the {\it l.h.s.} of Fig.~\ref{fig:abs_renyi_3fl} for tweaked Gaussian  points $X_{0.25, 20, \{5, 18\}}$. The corresponding {\sf metamorph} solutions are illustrated in Fig.~\ref{fig:abs_renyi_vertices}. This exercise, for the pion, has the peculiarity of including a zero gluon PDF as an acceptable solution. In the sense of R\'enyi entropy, it is a flat, uncertainty distribution, but it does not integrate to 1. Hence, it contributes with statistical noise,  slightly preventing more general conclusions, see Appendix~\ref{app:diffGlue} for discussion. 

\begin{wrapfigure}{R}{0.5\textwidth}
     \centering
     \includegraphics[width=0.85\linewidth]{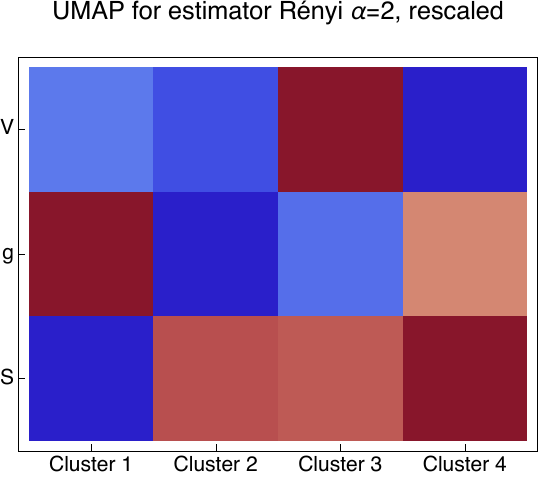}
\caption{Color matrix based on the value of the estimator $H_2^{\{V, g, S\}}(X)$ for UMAP-identified clusters displayed in the right panel of Fig.~\ref{fig:abs_renyi_vertices}.  The values of the average R\'enyi entropy are rescaled between $\left[0,1\right]$ (from dark blue to dark red) for improved visualization. To each cluster corresponds a different pattern in projected entropies.}
\label{fig:colormatrix_renyi}
\end{wrapfigure}

The solutions that are selected through the vertices correspond, in methodology, to the Fant\^omas pion PDF~\cite{Kotz:2023pbu}, for which we hand-picked the most diverse shapes. The latter were then  combined~\cite{Kotz:2025lio, Kotz:2025une} through the \texttt{mcgen} technique~\cite{Gao:2013bia}, for which each of the solution is associated to a distribution through its corresponding uncertainty in the Hessian formalism. Then, appealing to the de Finetti's theorem, we argued that exchangeable,  and independent and identically distributed distributions must find a common origin from an underlying truth, yet unknown, distribution.  The selection criteria described here completes the full methodology of model combination.\\

Yet another way to use R\'enyi entropy is as an estimator for clustering the solutions. Indeed, the coordinate vector can be used either in an in-house clustering prescription or in more established techniques requiring marker vectors. We pursued this exercise using the online embedding projector tool\footnote{\url{https://projector.tensorflow.org/}} with the UMAP (Uniform Manifold Approximation and Projection) dimensionality reduction algorithm. While UMAP is not a clustering algorithm {\it per se}, it leverages distances among nearest neighbors to define the relevant axes in the space of effective dimensions.
 For the pion case, dimensionality reduction is not strictly necessary 
(flavor space is three-dimensional: $v, g, S$), but examining the closest neighbors in the evaluated estimator ({\it vector}) provides useful information for classifying solution IDs ({\it metadata}). By using the R\'enyi entropy for all three flavors as coordinates of the ({\it vector}) estimator and the solution labels as ({\it metadata}), we identified four distinct regions, or clusters, with UMAP. These clusters are then mapped back into the actual $n_{\rm flavor}$-dimensional space, as illustrated in the right panel of Fig.~\ref{fig:abs_renyi_vertices}, along with the vertices of the Pareto hypersurfaces. 
 In this  case, each cluster contains at least one Pareto vertex, {\it i.e.}, it confirms that the Pareto vertices are representative of all clusters. Fig.~\ref{fig:colormatrix_renyi} illustrates
 average estimator patterns
 in \ren space for each identified cluster, confirming the difference between clusters, as well as a balance between smaller and larger entropies, on average, across flavors.

\subsection{Relative  criteria}
\label{sec:relative}

A common approach is to use divergences, which have an information-theoretic basis, and metrics, which are topological in nature, to classify and quantify distances between a pool of solutions and a reference baseline. On the information-theoretic side, the Kullback–Leibler (KL) divergence has been widely used both to quantify solutions~\cite{Carrazza:2015hva} and to profile PDFs~\cite{Hu:2025bla}.

Since the problem that we tackle in this paper is not necessarily characterized by a central value, 
evaluating relative measures requires comparing each solution with the others. It can become combinatorially expensive for a large number of trialed solutions. In the present exercise, only about $N_s=20$ solutions pass the pre-processing filters, rendering this cross-analysis manageable. We take advantage of this to show that the \ren entropy produces satisfactory results, thereby reducing the need for longer computations. 
This is crucial when applying selection criteria to problems with larger solution sets, such as the case of the proton PDF~\cite{Hou:2019efy,Ablat:2024muy}.\\

In this Section, we consider both an information-theoretic measure, the \ren divergence, and a metric measure, the Wasserstein distance, as a way to cross-check the results obtained with an absolute criterion.

\subsubsection{Information character -- R\'enyi divergence}

A natural extension of our framework for comparing relative shapes involves the \ren divergence, defined as,  
\begin{eqnarray}
    D_\alpha(p||q) = \frac{1}{\alpha-1} \ln \left( \sum_{i=1}^{n} \frac{p(x_i)^\alpha}{q(x_i)^{\alpha-1}} \right), \quad \alpha \neq 1
    \label{eq:renyi_div}
\end{eqnarray}
with $p$ the probability distribution from which $p(X)$ is drawn, and $Q$ for $q(X)$. For $\alpha=1$, $D_\alpha (p||q)$ is nothing else than the KL divergence, where $q$ plays the role of the reference density and $p$ comes from the pool of trials. \\

Here, nor $p$ or $q$ represent a baseline. In practice, we will take $\{p, q\}$ from the poll of solutions $\underline{f}=\{x f_k, k=1, \cdots, N_s\}$, and $\alpha=2$,
\begin{eqnarray}
    D_2(f_j||f_k) =  \ln \left( \sum_{i=1}^{n} \frac{\left(x_i\, f^j_{a/h}(x_i)/\sum_{l=1}^{n}x_j\, f^j_{a/h}(x_l)\right)^2}{\left(x_i\, f^k_{a/h}(x_i)/\sum_{l=1}^{n}x_j\, f^k_{a/h}(x_l)\right)} \right), \quad j \neq k \quad.
    \label{eq:renyi_div}
\end{eqnarray}
By definition, the \ren divergence is asymmetric, {\it i.e.}, $D_2(f_j||f_k)\neq D_2(f_k||f_j)$. The evaluation of all possible divergences involves $N_s (N_s-1)$ terms per flavor. For the example we are considering, the time involved for $N_s\sim 20$ is nearly 30 times longer than for the evaluation of the set of \ren entropies.\\

All divergences are  normalized to $D_2\in [0,1]$, and  then, again, considered as coordinates in the $n_{\rm flavor}$-dimensional space. Since we would like to select the most diverging pairs, we identify the $n_{\rm sol}=5$ points on the Pareto fronts with the highest norm as our best pairs -- they are all located on the roof. We take the set of the {\sf metamorphs} appearing in the vertices pairs as the desired set. This result is illustrated by the (long-dashed) mustard curves in Fig.~ \ref{fig:comp_Ren_div_W1}. For all flavors but the gluon\footnote{See Appendix \ref{app:diffGlue} for discussion on the gluon case.}, the outer limits of the selection based on the \ren divergence are close to those obtained through taking the vertices of the Pareto fronts of the \ren entropy space.

\subsubsection{Metric character -- Wasserstein distance}

The Wasserstein metric exploits the topology of the problem, and  measures the distance between two probability distributions $P$ and $Q$ on a metric space, characterized by a set $\Omega$ and a metric $d$. It evaluates the minimum cost of transforming one distribution into another, where the cost is a function of the distance between the two distributions -- it is akin to the work per unit mass for moving the masses. Hence, $W_p$ represents a ``horizontal distance" as opposed to the ``vertical distances," like $L^p$~\cite{santambrogio2015optimal}. As such, it captures the displacements of the curve along the horizontal axis, which, for functions with support on $[0,1]$ translates into shifts in the shape. It is a key tool in the context of  optimal transport theory~\cite{villani2009optimal, santambrogio2015optimal}.

For probability distributions $P$ and $Q$ on $(\Omega, d)$, the $p$-th Wasserstein distance  $W_p(P, Q)$ is defined as:
\begin{eqnarray}
  W_p(P, Q) = \left( \inf_{\gamma \in \Gamma(P, Q)} \int_{\Omega \times \Omega} d(x, y)^p \, d\gamma(x, y) \right)^{\frac{1}{p}},  
\end{eqnarray}
where \( \Gamma(P, Q) \) is the set of all joint distributions \( \gamma \) on \( \Omega \times \Omega \) with marginals \( P \) and \( Q \). $d\gamma(x,y)$ represents the hypersurface where the mass that is moved from $x$ to $y$ lies, and 
\begin{eqnarray}
    F_{X,Y}(x, y)&=& \int_{-\infty}^x \int_{-\infty}^y \, d\gamma(u, v) du dv\quad, 
\end{eqnarray}
is its joint cumulative distribution function. The measures of masses, from $x$ to $y$, are hence mapped by $d\gamma(x,y)$. $d(x, y)$ is the distance between those mass distributions, and the cost is expressed as $d(x, y)^p$. 
The infimum, $\inf$, is the greatest lower bound, here, of $\gamma \in \Gamma(P, Q)$.

\begin{wrapfigure}{r}{0.5\textwidth}
\centering
\includegraphics[width=0.9\linewidth]{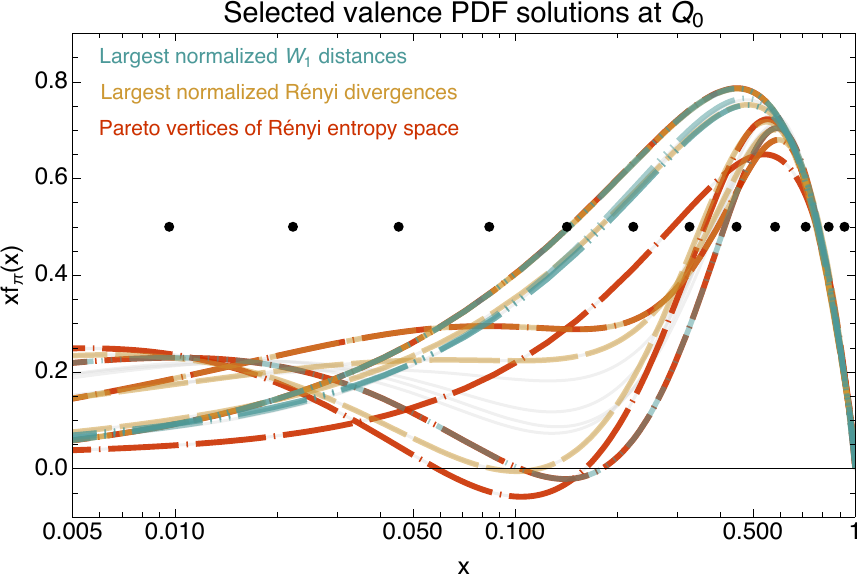}\\
\includegraphics[width=0.9\linewidth]{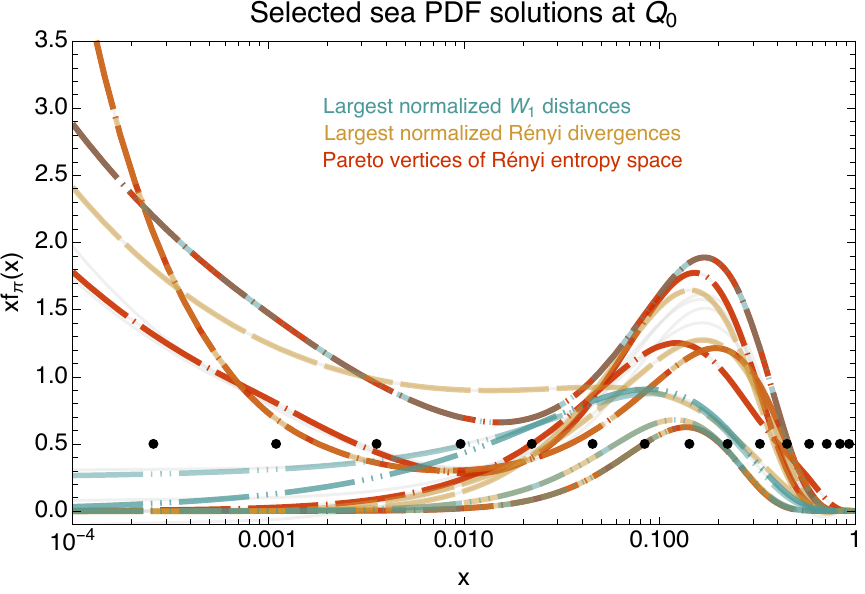}\\
 \includegraphics[width=0.9\linewidth]{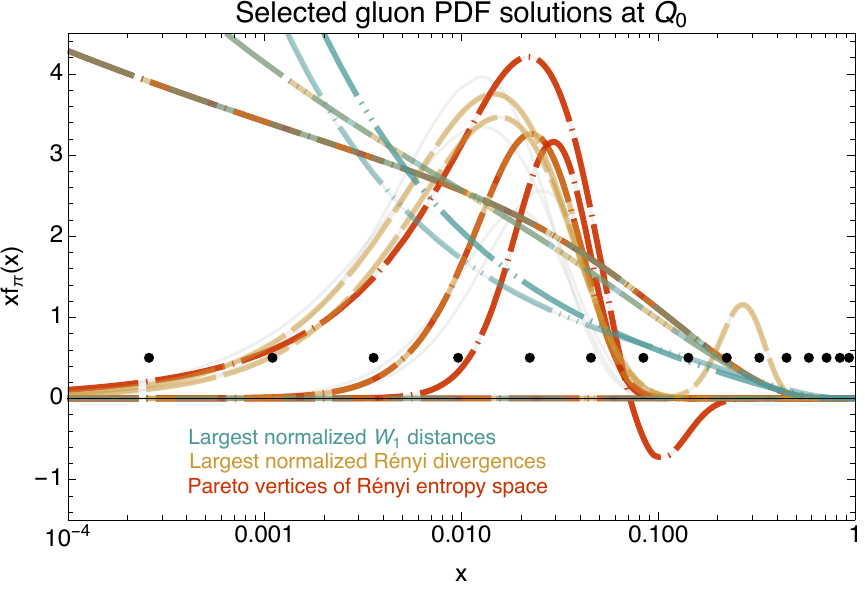}
\caption{Comparison of the selection obtained by the absolute criterion (dotted-dashed red), and two relative criteria -- \ren divergence (long dashed mustard) and Wasserstein distance (3 dots-dashed cyan).}
\label{fig:comp_Ren_div_W1}
\end{wrapfigure}

In one dimension and for the specific case of a translation $\gamma(x, y)= Q(x) \delta(y-T(x))$, the Wasserstein distance can be expressed as
\begin{eqnarray}
  W_p(P, Q) = \left( \int_{\Omega} |x-T(x)|^p\,  dQ(x)\right)^{\frac{1}{p} }.  
\end{eqnarray}

In practice, we will use projected PDFs, $x f(x)$. We choose $p=1$, known as the Earth Mover's Distance (EMD), in which case the flow of ``mass" can be recast in such a way as to write $W_1$ directly in terms of the cumulative distributions in $x$. As we will consider the discrete case, our final expression reads
\begin{eqnarray}
  &&W_1(f_j, f_k) \\
  &&=  \sum_{i=2}^n \left|\left( \sum_{l=2}^ix_l\, f_{a/h}^j (x_l)\right)-\left(\sum_{l=2}^i x_l\, f_{a/h}^k (x_l)\right)\right|\times \left| x_{i}-x_{i-1}\right| ,  \nonumber
\end{eqnarray}
where the discretization, in the Riemann sense, is accounted for through $| x_{i}-x_{i-1}|$. 
The Wasserstein distance is symmetric, hence it involves the combinatory of $(N_s, 2)$, reducing the CPU time {\it w.r.t.} the \ren divergence case by a factor $2$. Yet, it is still $15$ times longer than evaluating \ren entropy, for the $N_s\sim 20$ case. The selection of the most different shapes proceeds as for the \ren divergence. The $W_1$ distances act as coordinates in the $n_{\rm flavor}$-dimensional space, the Pareto fronts of the scattered points are evaluated, and the $n_{\rm sol}$ points with largest norm are picked.  

The resulting {\sf metamorphs} are shown in (3-dots dashed) cyan on Fig.~\ref{fig:comp_Ren_div_W1}. The selection highlights complementary aspects {\it w.r.t.} those obtained through information criteria, due to $W_1$ focusing on horizontal changes in the shapes.
Yet, the results for the valence and sea sectors are qualitatively consistent with those obtained through evaluating directly the \ren entropies as an absolute shape estimator. The gluon case shows differences between the selections, a peculiarity that comes from the presence of atypical shapes for PDFs. Yet, the results are acceptable, considering that the curves presented here will be accompanied by their respective error band, to be combined~\cite{Gao:2013bia,Kotz:2025lio}. This is further discussed in Appendix~\ref{app:diffGlue}.

While the case of collinear PDF seems too simple to resort to such topological metrics, the advantage lies in the generalization to higher dimensional problems. We note that the Wasserstein distance has been used in jet algorithms as a shape selection metric, in particular EMD~\cite{Komiske:2019fks, Komiske:2020qhg, Ba:2023hix}, due to its topological connection to the measure of mass distribution.

\section{Discussion}
\label{sec.discussion}

The objective of the present analysis is to provide an algorithm for shape selection through classification via estimators. The studied shapes correspond to the models for the pion PDFs, defined on $x\in [0,1]$. As shape estimator, \ren entropy provides a reliable {\it absolute} characterization of each curve separately. The CPU time that its evaluation requires is advantageous. 
The other two estimators that we have explored both  characterize shapes in pairs, {\it i.e.} they constitute {\it relative} estimators. The \ren divergence, with its information-theoretic character, and the Wasserstein distance, with its metric character, turn out to be complementary measures.

Once the estimator evaluated, it fills the $n_{\rm flavor}$-dimensional space with the coordinates of each PDF model. Now, at the selection step, the optimal distribution of the diverse shapes is nothing else than the Pareto fronts (Pareto front and worse front). In the case of the absolute estimator, we pick the vertices of both fronts as the compromised selection of most representative PDF models. This is justified by the expected forms of PDFs. As for the relative estimator, the $n_{\rm sol}$ models on the Pareto fronts with largest norm are chosen.
All three estimators lead to consistent and coherent results for sets of continuously changing shapes. The evaluation of \ren entropies is less computationally demanding, hence more advantageous.

Ideally, one would like to provide a full clustering algorithm. Instead, we show that clustering solutions for PDFs may not be necessary if the goal is the identify representative solutions. In \ren entropy space, the Pareto fronts supply the role played by a two-step selection, that is the clustering followed by the identification of the best solution in each cluster. This is true if the solutions have been filtered by their $\chi^2$ values prior to the selection procedure. However, \ren entropy makes a particularly suitable estimator for clustering. While we have not fully pursued this path, we have shown its combined usage to UMAP dimensionality-reduction method. \\

A few comments and caveats shall still be discussed before we draw our final conclusions.
\\

As \ren entropy estimates the amount of information, it differs from the measure of {\it wiggliness}, the kinetic energy, proposed in Ref.~\cite{Ball:2022uon} to tackle signs of overfitting, and that effectively accounts for the length of the curve. In this work, we did not attempt to characterize overfitting using information-theoretic criteria.\\

It may appear at a first glance that both Wasserstein and Pareto's approaches consider hypersurfaces, or surfaces whose continuous limit are hypersurfaces in the case of the Pareto fronts defined earlier, as the result of optimization. Yet, their nature and usage are completely different. Wasserstein distance maps hypersurfaces, it is of topological nature and characterizes relative distances. On the other hand,  Pareto orders discrete points to form an optimal front, which is used to define borders in some discrete parametric space.

In this analysis, we use Pareto fronts, not as an efficiency measure, but as a divider of coordinates in R\'enyi space. If the entropies were indeed a measure of optimization, {\it e.g.}~\cite{Gambhir:2025xim,GarciaCaffaro:2025gkm}, then the Pareto fronts would lead to an efficiency measure. However, we can interpret the Pareto floor as the optimally best known distributions --a balance between the smallest combination of $n_{\rm flavor}$ values, which correspond to more concentrated, less entropic, shapes-- and the Pareto roof as the optimally worse known distributions -- conversely, a balance between the largest combinations of $n_{\rm flavor}$ values, which correspond to flat, more entropic, shapes.\\

The selection of the most diverse shapes based on the relative estimators presents an extra layer of complexity {\it w.r.t.} that based on the absolute estimator. While it is easy to identify the largest differing pairs from the Pareto roof of the $n_{\rm flavor}$-dimensional space, some solutions may pair among each other to give small distances. Such pairs lie on the Pareto floor and they can easily be identified, and removed from the selection.\\

Another important aspect of the classification algorithms that we have proposed is the dependence on the choice for the vector $X$. 
In the above analysis, we have reported results for a specific choice for the evaluation of the various entropies, divergences and distances. We have explored the results with different choices of either $\alpha_x$ parameter, $n$ or even different spacing choices. While the results mildly vary with the choice of the $X$ vector, the qualitative output is not substantially affected. As a matter of fact, the choice of the $x$ span is part of the strategy to emphasize regions of expected spreads in parametrizations. The model for $X$ is captured through the ``hyperparameters" $(\alpha_x, n, \tau)$. 

In the same line of thought, the freedom to choose $X$ also allows to define partial \ren entropies, for example at large vs. small-$x$ values, a strategy that we have explored, yet not pursued, for one or various of the $n_{\rm flavor}$ dimensions.  We retain that this choice is an integral part of the definition of the shape estimator in the classification process.\\

The \ren entropy-based algorithm could be extended to the study of solutions for the proton PDF. In recent analyses of the CTEQ-TEA collaboration, the role of parametrization bias has been accounted for through the concept of tolerance, consisting in expanding the $\Delta \chi^2$ cut, which encompasses most solutions~\cite{Hou:2019efy}. On-going analyses aim to quantify the parametrization bias, an effort to which the present work sums up. For the proton, spaces are $6$-dimensional, at least. A preliminary usage of the \ren entropy as an estimator ({\it vector}) in the UMAP technique displays promising results for pattern classification~\cite{houston2025pdf4lhc}. On the other hand, given the properties of the Pareto fronts (see Appendix~\ref{app:pareto}), the extension of the algorithm for the \ren entropy selection could straightforwardly be extended to $n_{\rm flavor}=6$-dimensional spaces.

\section{Conclusions}
\label{sec:conclusions}

Parton Distribution Functions embody the probability of finding a parton with a momentum fraction $x$ of its hadron parent, at a given scale. Being highly non-perturbative objects, PDFs, as unknown functions of $x\in [0,1]$, are best determined through data-driven analyses within the theoretical framework of QCD. 
Many models can be designed that will lead to optimized solutions, or multiple distinct solutions to a single model can be found, in the Monte Carlo approach, that show different patterns. The question arises as to how these solutions should be interpreted

On the one hand, sampling over models, and hence generating more solutions, contributes to a specific case of epistemic uncertainty, namely model uncertainty~\cite{Courtoy:2022ocu, Kotz:2023pbu, Kotz:2025lio, Kotz:2025une}. Its contribution to the total uncertainty of PDFs is crucial for obtaining robust and accurate predictions. It is achieved via a combination method~\cite{Gao:2013bia, PDF4LHCWorkingGroup:2022cjn}, which produces a single set of PDFs. 
At present, the distributional nature of functional forms is not fully understood, although hints point toward the existence of higher parent distributions~\cite{Kotz:2025lio}. The pragmatic view-point on understanding the impact of the various possible solutions, that we adopt here, consists in identifying the most diverse solutions. Those solutions are regarded as $n_{\rm flavor}$ shapes, which we map onto estimators, so to classify and select the most diverse among them.

On the other hand, identifying patterns in PDF shapes in the Monte Carlo approach may also prove valuable, {\it e.g.} for assessing the role played by priors in the original bootstrapped distributions. 
\\

In this paper, we have proposed algorithms to address the problem of selecting shapes of PDFs. We have used both information-theoretic   and optimal-transport based estimators, which differ significantly in how they measure diversity. The former quantifies how much a distribution is structured (a measure of {\it bumpiness}), while the other reflects the cost of shifting the shapes. Furthermore, we classify the estimators as {\it absolute}, referring to single-shape measures, as opposed to {\it relative}, which involve pairwise comparisons.

We advocate for the use of the \ren entropy as a shape estimator for PDF solutions. It characterizes curves by a single number, hence we referred to \ren entropy as an {\it absolute} estimator. While the \ren entropy neither uniquely nor unambiguously determines a function, it appears to perform well in characterizing the space of solutions -- each of them corresponds to $n_{\rm flavor}$ coordinates in \ren space. The crucial step following that characterization consists in selecting the most diverse shapes, which encompass a large fraction of all possible solutions.  This is achieved identifying the Pareto floor and roof of the ensemble of coordinates in \ren space. The fronts lie inside the convex hull formed by the coordinates of all solutions, 
and the range of values spanned is determined by the vertices of the union of these fronts. Tested on the Fant\^omas pion PDFs, the results thus obtained are qualitatively similar to those presented in the original FantoPDFs~\cite{Kotz:2023pbu}. More interestingly, \ren entropy also proves to be a promising estimator for use in clustering algorithms. 
Both directions should be further explored in the context of proton PDF selection.

\section*{Acknowledgments}

AC thanks her CTEQ-TEA colleagues as well as the Fant\^omas team for fruitful discussions on model uncertainty and selection criteria. AC and AI are grateful to the PICO initiative Summer 2024 at UNAM during which this project started. AC is supported by the UNAM Grant No. DGAPA-PAPIIT IN102225. This work was performed in part at the Aspen Center for Physics, which is supported by a grant from the Simons Foundation (1161654, Troyer).

\appendix

\section{Tweaked Gaussian points}
\label{app:tweakGauss}

The $i$-th Gauss node/point, $x_i$, is the $i$-th root of $P_n$, the Legendre polynomial of degree $n$, with respective weights $w_i$ given by the formula
\begin{eqnarray}
    w_i&=& \frac{2}{(1-x_i^2)\, [P'_n(x_i)]^ 2}\nonumber
\end{eqnarray}

Using the Gauss-Legendre quadrature for functions with support in $x\in [0,1]$ as well as for the usual stretched/tweaked $x$ power, $x^{\alpha_x}$, the Gaussian points and weights are such that
\begin{eqnarray}
    \int_{}^1 f(x) dx\simeq \sum_{i=1}^n \, w'_i f(x'_i)\, \quad {\rm with}\quad P_n(x_i^{\prime\,\alpha_x})=0\quad.
\end{eqnarray}
We will refer to $X_{\alpha_x, n, \tau}$ for the vector of truncated tweaked Gaussian points obtained from $P_n(x^{\alpha_x})$. The truncation  will be specified by $\tau$.\

\section{Pareto fronts  as boundaries}
\label{app:pareto}

The properties of the Pareto fronts have been largely studied. The most relevant ones for our purposes can be found in  Emmerich~ \cite{Emmerich:2018} and Ehrgott~ \cite{Ehrgott:2005}. \\

Pareto fronts arise in optimization problems where more than one objective function/estimator must be considered simultaneously. The Pareto dominance in  such problems defines a strict {\it partial order} for the objective space $\mathbb{R}^n$, {\it i.e.} it is defined by componentwise order (inequalities) among solutions in the $n$-dimensional space. Following the definition of the Pareto floor, {\it i.e.} Pareto front, and roof, {\it i.e.} worse front, we summarize a few key properties.

A simple geometric property is that the union of floor and roof points lies inside the convex hull of all points,
\begin{eqnarray}
    P_{\rm floor}(\mathcal{Y})\cup P_{\rm roof}(\mathcal{Y})\subset\operatorname{conv}(\mathcal{Y})\quad,
\end{eqnarray}
with ${\cal Y}$ the set of points, and $\operatorname{conv}(\mathcal{Y})$ is the convex hull formed by all points in the objective space. Following~\cite{Ehrgott:2005}, in optimization problems, the argument is strengthened, {\it i.e.} the boundary points of the convex hull correspond to Pareto optimal points,
\begin{eqnarray}
    P_{\rm floor}(\mathcal{X})\cup P_{\rm roof}(\mathcal{X})\subseteq\operatorname{conv}(\mathcal{Y})\quad,
\end{eqnarray}
where ${\cal X}$ is the set of all possible solutions to the optimization problem taking ${\cal Y}=f(\cal{X})$. For example, given $x_1,x_2\in\mathcal{X}$, possible solutions to the optimization problem, we say that $x_1$ Pareto dominates $x_2$ if and only if $\underline{f}(x_2)\succ f(x_1)$, denoted $x_2\succ_{ f}x_1$. The symbol $\preceq$ is used to denote a partial order in $\mathbb{R}^k$, defined componentwise,
\begin{eqnarray}
        \underline{y}' \preceq \underline{y} \quad \Longleftrightarrow \quad y'_i \le y_i, \quad \forall i = 1, \dots, k
\end{eqnarray}
which is the definition of the Pareto front given in Eq.~(\ref{eq:pareto}).\\

Of relevance for the extension of our algorithm to higher dimensions is the following: the property that all solutions are bounded by both fronts, {\it i.e.}\\

\noindent $\mathcal{X}$ is \emph{bounded in the Pareto sense} by $P_{\rm floor}(\mathcal{X})\cup P_{\rm roof}(\mathcal{X})$
if there exist $y^{\min}, y^{\max} \in P_{\rm floor}(\mathcal{X})\cup P_{\rm roof}(\mathcal{X})$
such that $y^{\min} \preceq y \preceq y^{\max}$ for all $y \in \mathcal{X}$~\cite{Ehrgott:2005}.\\

\noindent That is, every element of $\mathcal{X}$ lies between its Pareto floor and roof limits with respect to the Pareto order, $\preceq$.

\section{The case of the gluon PDF of the pion}
\label{app:diffGlue}

\begin{wrapfigure}{r}{0.5\textwidth}
     \centering
     \includegraphics[width=0.95\linewidth]{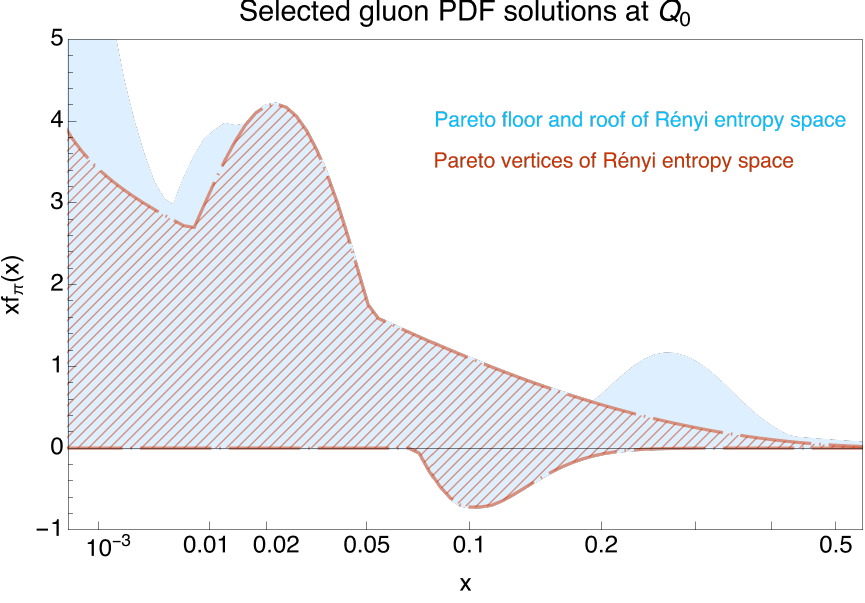}
\caption{Comparison of spanned solutions for the Pareto roof+floor vs. only vertices solutions for the gluon PDF of the pion. }
\label{fig:roflVstop_gl}
\end{wrapfigure}
The Fant\^omas PDF for the pion admits solutions for the gluon that include cases with a nearly zero contribution.  Motivated by Drell–Yan data~\cite{Kotz:2023pbu}, such solution yields phenomenologically interesting gluon momentum fractions, that are somewhat in tension with lattice QCD predictions.

Because its integral vanishes, this solution cannot be interpreted as a probability density. Its \ren entropy therefore arises purely from fluctuations in the gluon contribution. 
However, since \ren entropy, like many information criteria, characterizes only the shape of the distribution, not its magnitude, we expect the zero-gluon solution to be close to that of a flat distribution. Following Eq.~(\ref{eq:renyi_pract}), we obtain $H_2^{xf(x)=1}(X)=\ln{n}$.  The zero-gluon distribution gives $H_2^{xg(x)=0}(X)\lesssim\ln{n}$, which is sufficient to justify keeping this solution for our analysis.

On the other hand, the pion's gluon PDF is mainly determined by low-$x$ data from tagged deep-inelastic scattering data from a virtual pion target and,  through the DGLAP equations, by the better-constrained valence sector at very large-$x$. This leaves an extrapolation region around $x\sim 0.1$, where more atypical forms can emerge.  The bumpiest of the admitted solutions appear in this extrapolation region, although some are shifted in $x$. Because \ren entropy is insensitive to the $x$-position of the bump, the entropy of the function does not vary with the sliding of its maximum. Consequently, the peculiar shapes of the gluon PDF of the pion do not fully satisfy the simpler rule that the Pareto vertices should encompass all shapes  selected from both Pareto fronts.  This implies that, in this particular case, taking the vertices of the Pareto fronts is not sufficient, {\it i.e.} some of the steep solutions will be missed, as illustrated in Fig.~ \ref{fig:roflVstop_gl}. This issue is not expected to arise in the proton case, for which the extrapolation regions lie toward the end points.

\bibliography{distances}
\end{document}